\documentclass[12pt,a4paper]{article}
\usepackage[utf8]{inputenc}
\usepackage{import}
\usepackage{graphicx}
\usepackage[margin=1.5cm]{geometry}
\usepackage{amsmath,amssymb,mathrsfs,graphicx,indentfirst,amsfonts,listings}
\usepackage{float}
\usepackage{verbatim}
\usepackage{bbm}
\usepackage{enumerate}
\usepackage{mathtools}
\usepackage[linesnumbered,ruled]{algorithm2e}
\usepackage{xcolor}
\usepackage{caption}
\usepackage{multirow}
\usepackage{nameref,hyperref}
\usepackage{cite}

\makeatletter
\renewcommand{\@biblabel}[1]{\quad#1.}
\makeatother

\title{\Large
Should public health policy exempt cases with low viral load from isolation during an epidemic?: a modelling study 

}
\date{}
\author{Jiahao Diao
\thanks{School of Mathematics and Statistics, The University of Melbourne, Parkville, Victoria, Australia}\and Rebecca H. Chisholm\footnotemark[4] $^{, }$\thanks{Department of Mathematical and Physical Science, La Trobe University, Bundoora, Victoria, Australia}\and Nicholas Geard\thanks{School of Computing and Information Systems, The University of Melbourne, Parkville, Victoria, Australia}
\and James M. McCaw\footnotemark[1] $^{, }$ \thanks{Centre for Epidemiology and Biostatistics, Melbourne School of Population and Global Health, The University of Melbourne, Parkville, Victoria, Australia} $^{, }$ \thanks{Corresponding author: jamesm@unimelb.edu.au}}
\begin{document}
\maketitle
\begin{abstract}
	As demonstrated during the COVID-19 pandemic, non-pharmaceutical interventions, such as case isolation, are an important element of pandemic response. The overall impact of case isolation on epidemic dynamics depends on a number of factors, including the timing of isolation relative to the onset of contagiousness for each individual instructed to isolate by public health authorities. While there is an extensive literature examining the importance of minimising the delay from exposure to direction to isolate in determining the impact of case isolation policy, less is known about how underlying epidemic dynamics may also contribute to that impact. Empirical observation and modelling studies have shown that, as an epidemic progresses, the distribution of viral loads among cases changes systematically. In principle, this may allow for more targeted and efficient isolation strategies to be implemented. Here, we describe a multi-scale agent-based model developed to investigate how isolation strategies that account for cases viral loads could be incorporated into policy. We compare the impact and efficiency of isolation strategies in which all cases, regardless of their viral load, are required to isolate to strategies in which some cases may be exempt from isolation. Our findings show that, following the epidemic peak, the vast majority of cases identified with a low viral load are in the declining phase of their infection and so contribute less to overall contagiousness. This observation prompts the question about the potential public health value of discontinuing isolation for such individuals. Our numerical investigation of this `adaptive' strategy shows that exempting individuals with low viral loads from isolation following the epidemic peak leads to a modest increase in new infections. Surprisingly, it also leads to a \emph{drop} in efficiency, as measured by the average number of infections averted per isolated case. Our findings therefore suggest caution in adopting such flexible or adaptive isolation policies. Our multi-scale modelling framework is sufficiently flexible to enable extensive numerical evaluation of more complex isolation strategies that incorporate more disease-specific biological and epidemiological features, supporting the development and evaluation of future public health pandemic response plans.
\end{abstract}

\section{Introduction}
The recent COVID-19 pandemic demonstrated the vital role that non-pharmaceutical interventions (NPIs) play in mitigating the impacts of acute viral respiratory epidemics~\cite{gozzi2021importance,lai2020effect,spinelli2021importance}. In particular, isolation of cases, as part of test-trace-isolate-quarantine policies, was a key part of the public health response, slowing transmission of the virus and reducing the impact of the pandemic on society \cite{kucharski2020effectiveness, shearer2024estimating}. While widespread isolation of all cases can effectively reduce transmission of the virus and consequent impact on hospitalisations and deaths, it may also lead to substantial unintended health (e.g., mental health) and economic (e.g., interrupted employment) costs\cite{brodeur2021literature,talevi2020covid,ma2020effectiveness,ayouni2021effective,bargain2020trust}. In preparing for future pandemics, it is desirable to design NPIs that can reduce transmission while minimising unintended consequences.

One possible approach to minimising the social and economic costs associated with isolation is to make use of routinely available data on cases that may allow for more targeted application of isolation. In many jurisdictions, cases detected by qt-PCR also have an associated PCR cycle threshold (Ct) value reported to health authorities. The Ct value provides a quantitative measure of the viral load at the time of testing of each case \cite{rao2020narrative,finks2023clinical}. Given that an individual's contagiousness is likely an increasing function of their viral load---an assumption for which there is a growing body of empirical evidence \cite{marc2021quantifying,marks2021transmission}---it is therefore reasonable to consider whether those with low viral loads (high Ct values) could be \emph{exempt} from a requirement to isolate, because their lower contagiousness means they are unlikely to infect other people. However, a high Ct value (low viral load) could be measured at the end of an infection (where contagiousness is decreasing) or at an early phase of an infection (where contagiousness is increasing). Failing to isolate cases with high Ct values and increasing contagiousness could reduce the effectiveness of a more targeted isolation policy. Hay et al.~\cite{hay2021estimating} showed that the distribution of Ct values from cases varies over the course of an epidemic wave, suggesting that the proportion of cases with high Ct values and increasing contagiousness is also likely to change during an epidemic.  Therefore, the implementation of a more targeted isolation policy may have varying effectiveness at reducing infections during different phases of an epidemic.

Here, we describe a multi-scale agent-based epidemic model in which individual-level viral infection kinetics are resolved, and use it to explore the hypothesis that exempting low viral load cases from isolation during an outbreak will have minimal effect on the final size of an outbreak, while reducing the overall number of people who are required to isolate. We use the model to quantify the dynamic viral load distribution of agents throughout an unmitigated epidemic before using it to evaluate and compare several isolation strategies in which the decision to isolate is based upon viral load. We distill the problem down to its simplest form by considering the transmission of an SIR-like pathogen in which each infected individual's viral load is modelled using the target cell--infectious cell-virus (TIV) equations. An individual's contagiousness is specified to be a sigmoidal function of their viral load. We account for test sensitivity as a function of viral load, and consider scenarios in which only one, or multiple tests, are available for individuals during their symptomatic illness. We evaluate isolation management strategies in which all individuals who test positive to infection are isolated, and alternative strategies in which those with a low viral load may be exempt from isolation. We compare isolation strategies in terms of the reduction in the final size of the epidemic compared to a baseline with no isolation, and how many individuals were isolated to achieve that reduction in overall infections.

\section{Methods}
To investigate the value and consequences of alternative isolation strategies in which an individual's viral load determines if they are required to isolate, we developed a multi-scale agent-based transmission model, described in Section~\ref{ssec:multiscale}. The testing--isolation strategies are described in Section~\ref{ssec:testingisolation} and our evaluation framework is described in  Section~\ref{ssec:evaluation}.

\subsection{Multi-scale model}
\label{ssec:multiscale}

We consider a fixed population of $N$ individuals, with no births or deaths occurring during the simulation period. Each individual in the population is represented as an agent  $a_k$, where $k\in\{1,2,\cdots,N\}$. At each time step $t\in \{0,t_{\rm int},\dots, t_{\rm end}\}$, where $t_{\rm int}$ is the time step and $t_{\rm end}$ is the maximum simulation time, each agent is classified into one of three mutually exclusive sets: $\hat{S}(t)$ for susceptible, $\hat{I}(t)$ for infected and infectious, and $\hat{R}(t)$ for those who are removed and no longer infectious. The respective cardinalities of these sets are denoted by $S(t)$, $I(t)$, and $R(t)$, representing the total number of susceptible, infectious and recovered agents at time $t$.

The occurrence of infection events, in which a susceptible agent is reclassified as an infectious agent, are a function of the contact rate between agents and the probability of infection given contact between a susceptible and infectious agent. This probability depends on the within-host state of the infectious agent, as described below. Recovery events, in which an infectious agent is reclassified as removed, are also determined based on the within-host state of the infectious agent. 

Both an infectious agent's contagiousness and their time of recovery are functions of their time since infection, $\tau$, and in particular, their viral load. An agent's viral load is determined by a within-host scale compartmental model for their infection kinetics. In this sense, our model (without testing and isolation) is an agent-based implementation of the general model for an epidemic as introduced by Kermack and McKendrick~\cite{kermack1927contribution} and of which the classic SIR model is a special case in which an agent's contagiousness is constant and their time to recovery is exponentially distributed.

At the within-host scale, we assume each agent's viral kinetics are modeled by a target-cell limited model, that is the classic TIV model of viral dynamics \cite{nowak2000virus}:
\begin{eqnarray}\label{eq:TIV}
	\frac{dT_k}{d\tau}&=&-\alpha T_kV_k,\nonumber\\
	\frac{dL_k}{d\tau}&=&\alpha T_kV_k-\delta_L L_k,\nonumber\\
	\frac{dV_k}{d\tau}&=&pL_k-cV_k,
\end{eqnarray}
where $T_k$ is the number of target cells, $L_k$ is the number of infected cells and $V_k$ is the viral load of agent $a_k$. Note that we have used $L_k$ to denote infected cells simply to avoid ambiguity with the set of infectious individuals $\hat{I}$. The parameters in the TIV model are the infectiousness of the virus when interacting with target cells ($\alpha$), the rate of removal of infected cells ($\delta_L$), the rate of production of free virus by infected cells ($p$) and the rate of removal of free virus ($c$), as well as the initial conditions for each of the state variables, which we take to be $\{T_k=T_0, L_k=0, V_k=V_0\}$ with $T_0$ and $V_0$ positive constants. In principle, agents may have different values for each of the parameters of the within-host model, although for simplicity we take common values for all parameters (from~\cite{baccam2006kinetics}) except for $log_{10} V_0$, which for each agent is drawn from a uniform distribution $U(0,2)$ to introduce some (minor) variation in the within-host kinetics across infected agents.

We define the contagiousness of an infectious agent (i.e., those $a_k\in\hat{I}$) to be the probability of infection given contact with a susceptible individual. We model contagiousness $\tau$ units of time since the time of infection, denoted by $\beta_k(\tau)$, as a sigmoidal function of the agent's viral load $V_k(\tau)$:
\begin{eqnarray}\label{eq:infectiousness}
	\beta_k(\tau)=\frac{\beta_{max}V_k(\tau)^\xi}{V_k(\tau)^\xi+V_{50}^\xi},
\end{eqnarray}
where $\beta_{max}$ is the maximum contagiousness of the infected agent, $V_{50}$ is the viral load at which contagiousness is half-maximal, and $\xi$ is a slope parameter characterising the steepness of the transition from low to high contagiousness at $V_k (\tau) = V_{50}$. Our decision to adopt a sigmoidal function to map from viral load to contagiousness is common within the literature \cite{hart2020theoretical,henriques2022modelling,harris2023correlation} and has some experimental support, for example in~\cite{handel2015crossing,saini2012ultra}. An example trajectory of an infectious agent's contagiousness through time, given by Equation~\ref{eq:infectiousness}, is shown in \nameref{S1_Fig}.

By construction, use of the simple $TIV$ model guarantees that an agent's viral load decays to zero as $\tau\to\infty$~\cite{cao2016role}. Noting that for small $V_k(\tau)$, $\beta_k(\tau)$ is also (very) small, we reassign an agent from being infectious to removed once $V_k(\tau) < \epsilon$ and $dV_K/d\tau < 0$, with $\epsilon = 10^{-2}$. The numerical value for $\epsilon$ is essentially arbitrary, as long as it is sufficiently small. Note that the reassignment of an agent as removed (transferring them from the set $\hat{I}$ to the set $\hat{R}$) is primarily for accounting convenience, with a negligible impact on the transmission dynamics.

We define the changes in the number of susceptible, infected, and recovered agents from time $t-t_{\rm int}$ to time $t$ as $\Delta S, \Delta I, \Delta R$, given by:
\begin{eqnarray}
	\label{eq:transiton_details}
	\Delta S &=& S(t)-S(t-t_{\rm int}) = \sum_{a_k\in \hat{I}(t)} \sum_{i= 1}^{n_k}\mathbbm{1}_{p\leq \beta_k(\tau)},\nonumber\\   
	\Delta I &=& I(t) - I(t-t_{\rm int}) = \Delta S - \Delta R,\nonumber\\
	\Delta R &=& R(t) - R(t-t_{\rm int}) = \sum_{a_k\in \hat{I}(t)} \mathbbm{1}_{V_{k}(\tau)\leq \epsilon},
\end{eqnarray}
where $N_k(t)\sim \text{Poiss}(\lambda t_{\rm int})$, $\lambda$ is the mean number of contacts an infected agent makes with all other agents per day, $n_k \sim \text{Bin}(N_{k}(t),\frac{S(t)}{N-1})$ is the number those contacts that are susceptible, and $p\sim U(0,1)$. Algorithm~\ref{al:algorthim} describes our computational implementation of the multi-scale agent-based model.

\begin{algorithm}[t!]
	\caption{Multi-scale agent-based model algorithm}
	\label{al:algorthim}
	\textbf{Input}: Agents $a_k,k\in\{1,2,\dots,N\}$, $\hat{S}(0),\hat{I}(0),\hat{R}(0),\hat{Q}(0),S(0),I(0),R(0)$, $Q^{\rm{cml}}(0)$;\\
	\textbf{Output}: Agents $a_k,k\in\{1,2,\dots,N\}$, $\hat{S}(t_{\rm{end}}),\hat{I}(t_{\rm{end}}),\hat{R}(t_{\rm{end}}),\hat{Q}(t_{\rm{end}}),S(t_{\rm{end}})$, $I(t_{\rm{end}}),R(t_{\rm{end}}),Q^{\rm{cml}}(t_{\rm{end}})$;\\
	\textbf{Initialisation:} Set initial within-host states $T_k(0), L_k(0), V_k(0)$ for $\forall a_k\in\hat{I}(0)$ and infection times $t^I_k=0$;\\ 
	\ForEach {$a_k\in \hat{I}(0)$}
	{Simulate the TIV model with ($\log_{10} V_{k}(0)\sim U(0,2)$);\\
		Set $T_k, L_k, V_k$ through $\tau$;\\
		Set $D_k$ = minimum $\tau$ where $V_k(\tau)\leq\epsilon$;\\
		Set $t^R_{k}=t^I_{k}+D_k$;}
	
	\While{$I(t)\neq 0$ $\&\&$ $t< t_{\rm end}$}{    
		\ForEach{$t\in\{0,t_{\rm int},\hdots,t_{\rm end}-t_{\rm{int}}\}$}{
			Set $\hat{I}_{\rm new}=\hat{R}_{\rm new}=\hat{Q}_{\rm new}=\hat{Q}_{\rm leave}=\emptyset$;\\
			\If{\rm testing-isolation strategies}{Apply Algorithm~\ref{al:algorithm2}.}

			\ForEach{$a_k\in \hat{I}(t)$}{
				\uIf{$t\leq t^R_{k}$}{
					Generate number of contacts 
					$N_{k}(t)\sim \text{Poiss}(\lambda t_{\rm int})$;\\
					Generate number of susceptible contacts	
					$n_{k}(t) \sim \text{Bin}(N_{k}(t), \frac{S(t)}{N-1})$;\\
					Set $\hat{n}_k(t)$ = a simple random sample of $\hat{S}(t)$ of size $n_{k}(t)$;\\
					
					Set $\tau = t-t^I_{k} $;\\

					\ForEach{$a_j\in \hat{n}_k(t)$;}{
						
						\If{$p\leq \beta_k(\tau)$, where $p\sim U(0,1)$}{
							
							Set $\hat{I}_{\rm new}=\hat{I}_{\rm new}\cup a_j$;\\
							
							Set $t^I_{j}=t$;
							
							Simulate the TIV model with ($\log_{10} V_{j}(0)\sim U(0,2)$);\\
							
							Set $T_j, L_j, V_{j}$ through $\tau$;
							
							Set $D_j$ = minimum $\tau$ where $V_{j}(\tau)\leq\epsilon$;
							
							Set $t^R_{j}=t^I_{j}+D_j$;
						}
					}
					
				}
				\Else{
					Set $\hat{R}_{\rm new}=\hat{R}_{\rm new}\cup a_k$;}
			}
			Set $\hat{S}(t+t_{\rm int})=\hat{S}(t)\setminus \hat{I}_{\rm new}$;\\
			Set $\hat{I}(t+t_{\rm int})=\left(\hat{I}(t)\setminus(\hat{R}_{\rm new}\cup \hat{Q}_{\rm{new}})\right)\cup \hat{I}_{\rm new}$;\\
			Set $\hat{R}(t+t_{\rm int})=\hat{R}(t)\cup \hat{R}_{\rm new}$;\\
			Set $\hat{Q}(t+t_{\rm int})=(\hat{Q}(t)\setminus \hat{Q}_{\rm leave})\cup \hat{Q}_{\rm new}$;\\
			Calculate $S(t+t_{\rm int}),\ I(t+t_{\rm int}),\ R(t+t_{\rm int})$;\\
			Calculate $Q^{\rm{cml}}(t+t_{int})=Q^{\rm{cml}}(t)+|\hat{Q}_{\rm{new}}|$;
		}
	}
\end{algorithm}

\begin{figure}
	\centering
	\includegraphics[width = 0.9\linewidth]{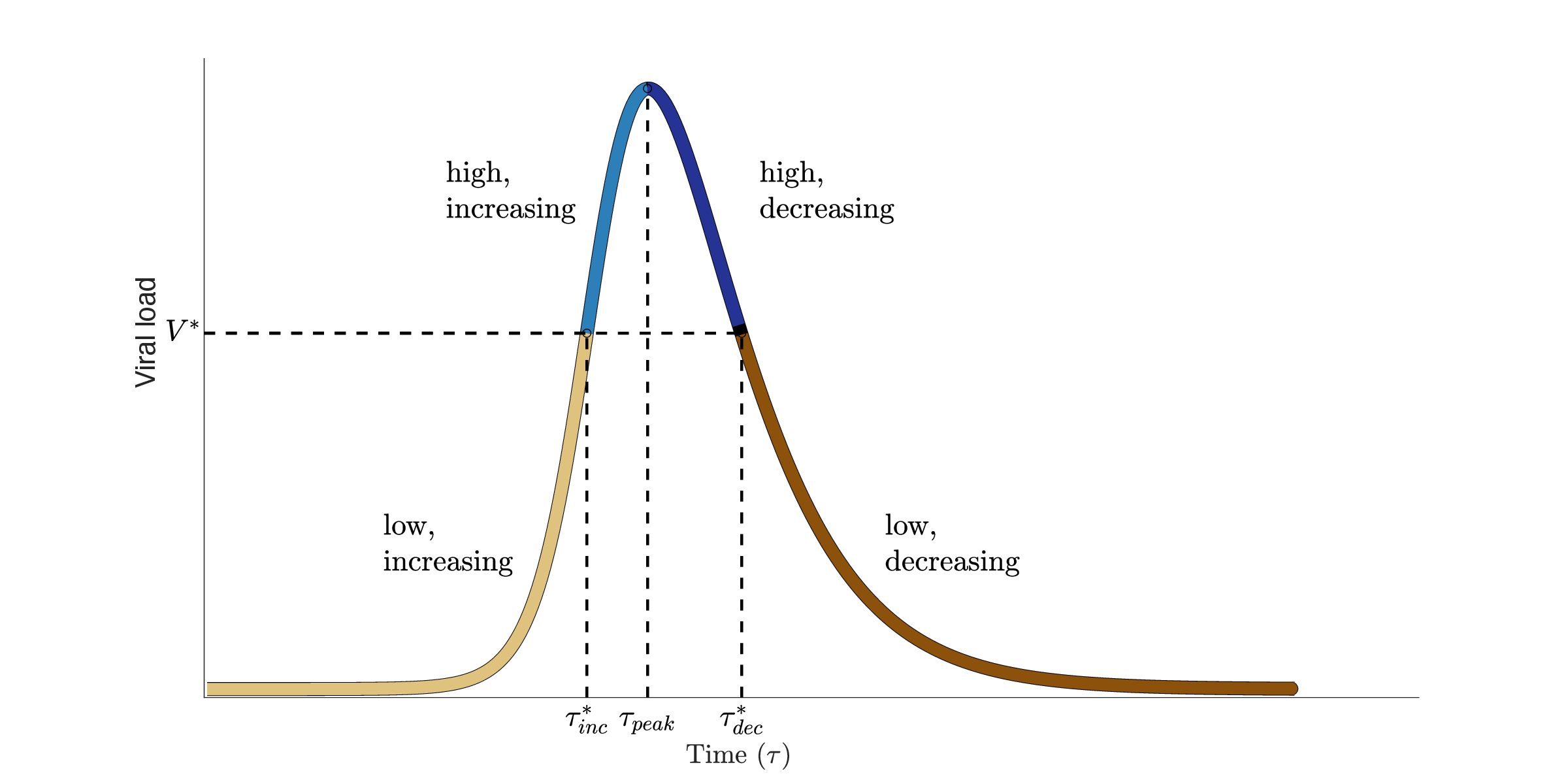}
	\caption{{\bf The four phases of infection for each agent}. The viral load kinetics for each agent are divided into four distinct phases: low increasing, high increasing, high decreasing and low decreasing. The threshold defining the transition from low to high viral load is denoted by $V^*$, and the three times ($\tau_\text{inc}^*$, $\tau_\text{peak}$, and $\tau_\text{dec}^*$) define the times of transition between the four phases.}
	\label{fig:basic}
\end{figure}

While an agent's viral load over time is a continuous trajectory, for convenience in defining and evaluating our alternative isolation strategies, we divide the infected period of each agent into four distinct phases, based on whether an agent's viral load is low or high, and whether it is increasing or decreasing (Figure~\ref{fig:basic}). The threshold viral load for classifying an agent as having a low or high viral load is denoted by $V^*$, and for the purposes of this paper, we choose this value such that the agent's contagiousness $\beta = \frac23 \beta_{max}$. Combined with the sign of $dV/dt$, we compute the times of transition between the phases: $\tau_\text{inc}^*$, $\tau_\text{peak}$, and $\tau_\text{dec}^*$.

With this classification for each agent, we can divide the set of infected agents into the following four sets at any time $t$ in the simulation:
\begin{itemize}
	\item {\bf $\hat{I}_{V_\ell}^\text{inc}(t)$}:= $\big\{\forall a_k\in\hat{I}(t)\ |\ V_k(\tau)<V^*\ \text{and}\ 0 \leq \tau < \tau_\text{inc}^* \big\}$;
	\item {\bf $\hat{I}_{V_h}^\text{inc}(t)$}:= $\big\{\forall a_k\in\hat{I}(t)\ |\ V_k(\tau)\ge V^*\ \text{and}\ \tau_\text{inc}^* \leq \tau < \tau_\text{peak} \big\}$;
	\item {\bf $\hat{I}_{V_h}^\text{dec}(t)$}:= $\big\{\forall a_k\in\hat{I}(t)\ |\ V_k(\tau)\ge V^*\ \text{and}\ \tau_\text{peak} \leq \tau < \tau_\text{dec}^* \big\}$;
	\item {\bf $\hat{I}_{V_\ell}^\text{dec}(t)$}:= $\big\{\forall a_k\in\hat{I}(t)\ |\ V_k(\tau)<V^*\ \text{and}\ \tau \geq \tau_\text{dec}^* \big\}$.
\end{itemize}

\subsection{Testing--isolation strategies}
\label{ssec:testingisolation}

In order for an infected individual to be isolated, they must be identified through surveillance (i.e., be recorded as a case). We only consider case detection due to symptom-based testing and do not model contact tracing, testing or quarantine of contacts, nor other aspects of surveillance such as mass screening. In our model, the probability that an infected agent is identified as a case depends on the presence of symptoms, the probability $P(\rm Tested)$ per unit time of seeking a test (symptom-dependent), and the sensitivity of the test, which is assumed to vary with viral load.


We model test positivity given infection as linear in the viral load:
\begin{eqnarray}\label{testpositive}
	P(\textrm{Positive}|\textrm{Tested \& Infected}) = \sigma_\text{min} + (\sigma_\text{max}-\sigma_\text{min})\times \frac{V(\tau)-\epsilon}{\max(V)-\epsilon}.
\end{eqnarray}
where $\sigma_\text{min}$ is the minimum test sensitivity and $\sigma_\text{max}$  is the maximum test sensitivity.
Figure \nameref{S2_Fig} presents the probability (per day) for an infected agent to be tested, and the probability of that test being positive as a function of the time since infection, for the two alternative model parameterisations (described below). 

We consider four strategies for isolation of cases (i.e., test-positive individuals), applied from very early in the epidemic (taken as day $t=5$~days) through to the end of the epidemic when no further infections occur (this is guaranteed to occur in finite time as we model stochastic SIR-dynamics with no susceptible replenishment in a finite fixed-size population):
\begin{itemize}\label{Strategy}  
	\item {\bf No isolation}: No infected agents are isolated;
	\item {\bf Isolate all}: All infected agents that tested positive are isolated;
	\item {\bf Isolate high}: Only infected agents  that have a high viral load at the time of testing (i.e., $\forall a_k\in \hat{I}_{V_h}^\text{inc}(t)\cup \hat{I}_{V_h}^\text{dec}(t)$) that tested positive are isolated; 
	\item {\bf Adaptive$_n$}: Begin with the \emph{Isolate all} strategy, before switching to the \emph{Isolate high} strategy $n$-weeks following the peak day (defined as the time of peak prevalence under the \emph{Isolate all} strategy), with ${n \in \{-5,-3,-1,0,+1,+3,+5\}}$. Note that $n<0$ indicates a switch prior to the peak day, and $n>0$ indicates a switch following the peak day.
\end{itemize}
Infected agents identified as cases and subject to isolation (strategy dependent) are assumed to be isolated immediately and for a fixed duration $d_\text{iso}$, and do not contribute to the force of infection in the population (i.e., their contagiousness is set to zero) during their isolation period. If an agent is released from isolation while still classified as infectious, they will once again contribute to the force of infection. 

For each of these strategies, we consider alternative testing scenarios in which each agent may test only once, or in which multiple (unlimited) tests may be conducted. Under the single test strategy, each infected agent has only one opportunity for testing (regardless of test outcome).  Therefore, an infectious agent with a low increasing viral load, if tested and positive (i.e., identified as a case), will not be isolated under the \emph{Isolate high} strategy, even though they will soon become highly contagious (entering their high increasing followed by high decreasing infection phase). Under a multiple testing scenario, such an agent could test positive later in their infection and be isolated. The same outcomes for an agent can potentially play out under the \emph{Adapative}$_n$ strategy at certain times in the epidemic, after the switch to only isolating cases with a high viral load has been made.

We denote by $\hat{Q}_s(t)$ the set of agents that are isolated at time $t$ under strategy \\
$s\in\{\emph{No isolation}, \emph{Isolate all}, \emph{Isolate high}, \emph{Adaptive}_n \}$. Isolated agents are reassigned to the recovered class in the same way as infected agents (i.e., once their viral load drops below the threshold $\epsilon$.) For convenience, we also record the cumulative number of isolated agents $Q_s^\text{cml}(t)$ up to time $t$.  Algorithm~\ref{al:algorithm2} describes the computational implementation of testing and isolation under the different isolation strategies.

\begin{algorithm}[t!]
	\caption{Algorithm of testing-isolation strategies}
	\label{al:algorithm2}
	\textbf{Input}: $\hat{I}_{\rm new}=\hat{R}_{\rm new}=\hat{Q}_{\rm new}=\hat{Q}_{\rm leave}=\emptyset$, and $d_{\rm{iso}}$: a fixed duration of isolation;\\
	\textbf{Output}:$\hat{I}_{\rm{new}},\hat{R}_{\rm{new}},\hat{Q}_{\rm{new}},\hat{Q}_{\rm{leave}}$;\\
	\ForEach{$a_k\in \hat{I}(t)$}
	{\If{${\rm AlreadyTested}_{a_k}=0$}
		{\If{$p_t\leq P(\rm Tested)$, where $p_t\sim U(0,1)$}{
				\If{$p_p\leq P({\rm Positive}\ | \ {\rm Tested})$, where $p_p\sim U(0,1)$}{
					\uIf{\rm strategies $=$ IsolateHigh or Adaptive}{
						\If{$\beta_{k}(\tau)\geq \frac{2}{3}\beta_{max}$}
						{Set $\hat{Q}_{\rm new}=\hat{Q}_{\rm new}\cup a_k$;\\
							Set $t_{k}^{\text{Isolate}}=t+d_{\rm{iso}}$;\\        
					}}
					{\uElseIf{\rm strategies $=$ IsolateAll}{
							Set $\hat{Q}_{\rm new}=\hat{Q}_{\rm new}\cup a_k$;\\
							Set $t_{k}^{\text{Isolate}}=t+d_{\rm{iso}}$;\\      
					}}
					\Else{\textbf{Continue}}
				}
				\If{\rm SingleTest}
				{$\text{AlreadyTested}_{a_k}=1$;}
			}
		}   
	}
	\ForEach{$a_k\in \hat{Q}(t)$}{
		\uIf{$t>t_{k}^R$}{
			Set $\hat{Q}_{\rm leave}=\hat{Q}_{\rm leave}\cup a_k$;\\
			Set $\hat{R}_{\rm new}=\hat{R}_{\rm new}\cup a_k$;
		}
		\Else{\If{$t>t_{k}^{\text{Isolate}}$}{
				Set $\hat{Q}_{\rm leave}=\hat{Q}_{\rm leave}\cup a_k$;\\
				Set $\hat{I}_{\text{new}}=\hat{I}_{\text{new}}\cup a_k$;
		}}
	}
\end{algorithm}

\subsection{Strategy evaluation}
\label{ssec:evaluation}
To evaluate the potential public health value of alternative isolation strategies---accounting for the trade-off between reducing transmission and the cost of restrictions on individuals---we introduce metrics that quantify the effectiveness and efficiency of isolation.

We define the effectiveness of an isolation strategy $s$ as the proportionate reduction in cumulative infections under strategy $s$ compared to the baseline with no isolation:
\begin{eqnarray}\label{eq:Effectiveness}
	\varphi_{s} &=& \lim_{t\to\infty} \left(1-\frac{C_s(t)}{C_\text{No isolation}(t)}\right),
\end{eqnarray}
where $C_s(t) = N-S_s(t)$ is the cumulative number of infections under strategy $s$. Under this definition, $\varphi_s\to 1$ indicates a dramatic (complete) reduction in cumulative infections compared to with no isolation, and $\varphi_s\to 0$ indicates negligible (no) reduction in cumulative infections compared to the baseline with no isolation.

When comparing scenarios, we initialise each simulation with the same random seed so that sample paths only differ after the relevant strategy implementation time.

As a measure of efficiency for an isolation strategy $s$, we measure the number of infections averted compared to the baseline with no isolation ($C_\text{No isolation}(t)-C_s(t)$) per isolated person ($Q^\text{cml}_s(t)$):
\begin{eqnarray}\label{eq:Efficiency}
	\text{E}_s &=& \lim_{t\to\infty} \left(\frac{C_\text{No isolation}(t)-C_s(t)}{Q^\text{cml}_s(t)}\right).
\end{eqnarray}

Finally, to evaluate the performance of the \emph{Adaptive}$_n$ strategy in more detail, we introduce an additional metric, the relative cost, $\psi_n$ for the \emph{Adaptive}$_n$ strategy. At best, the \emph{Adaptive}$_n$ strategy should perform as well as the \emph{Isolate all} strategy and at worst, perform as poorly as the \emph{Isolate high} strategy. We define $\psi_n$ as the proportionate loss of benefit (i.e., the relative cost) of the \emph{Adaptive}$_n$ strategy between these two extremes:
\begin{eqnarray}\label{eq:CompareSwitch}
	\psi_n &=& \lim_{t\to\infty} \left( \frac{C_{\text{Adaptive}_n}(t)-C_\text{Isolate all}(t)}{C_\text{Isolate high}(t)-C_\text{Isolate all}(t)}\right).
\end{eqnarray}

Accordingly, $\psi_n \to 0$ indicates that the effectiveness of the \emph{Adaptive}$_n$ strategy is the same as for the \emph{Isolate all} strategy (i.e., a cost of zero), while $\psi_n \to 1$ indicates that the \emph{Adaptive}$_n$ strategy has lost (on average) all of the benefit of having isolated cases with a low viral load at the time of detection, with effectiveness degraded (on average) to the same value as for the \emph{Isolate high} strategy.

\subsection{Model parameterisations}
We consider two different parameterisations of the within-host TIV model: (1) an ``influenza like'' parameterisation, with parameter values drawn from \cite{baccam2006kinetics}; and (2) a ``SARS-CoV-2 like" parametersiation, since SARS-CoV-2 has notably slower viral kinetics compared to influenza \cite{hernandez2020host}.  

Our choice of testing probabilities and test sensitivity are specific to each model parameterisation.  Under the ``influenza like'' parameterisation, we assume a two day incubation period, consistent with a range of empirical estimates for influenza \cite{bell2006non,canini2016heterogeneous,carrat2008time,lau2010viral}. Prior to symptom onset ($0\leq \tau < 2$) we assume there is a 20\% chance per day of seeking a test (i.e., $P(\rm Tested)=0.2$), while following symptom onset ($\tau \geq 2$) we assume this increases to an 80\% chance per day (i.e., $P(\rm Tested)=0.8$).  The minimum and maximum test sensitivities are set to $0.5$ and $0.8$, respectively~\cite{trombetta2018rapid,su2024performance}. For the ``SARS-CoV-2 like'' parameterisation, we assume a 3 day incubation period~\cite{puhach2023sars}. Prior to symptom onset ($0\leq \tau < 3$) we assume there is a 10\% chance per day of seeking a test (i.e., $P(\rm Tested)=0.1$), while following symptom onset ($\tau \geq 3$) we assume this increases to a 50\% chance per day (i.e., $P(\rm Tested)=0.5$). The minimum and maximum test sensitivities are set to $0.5$ and $0.8$, respectively~\cite{chu2022comparison,wells2022comparative}.

All outputs presented in the main manuscript were generated with the ``influenza like'' model parameterisation.  In the Supplementary Material, we present analagous outputs under the ``SARS-CoV-2" model parameterisation. 

For all scenarios presented (including those in the Supplementary Material), we consider a population of size $N=20,000$, with $\beta_{max} = 0.6$, initial conditions ${S(0)=N-10}$, ${I(0)=}10$, ${R(0)=0}$, and with all other parameters specified as per the Methods and summarised in Table~\ref{tab:table1}.  Our choice for $\beta_{max}$ results in an epidemiological-scale basic reproduction number of $\mathcal{R}_0 \approx 2.12$. With these parameter choices, initial conditions, and without any intervention, it is unlikely for stochastic extinction to occur during the early phase of an epidemic. The code used for this work is available at: https://github.com/Effectiveness-of-isolating-low-viral-load-Multi-scale-model.

\begin{table}[t!]
	\scriptsize    
	\centering
	\begin{tabular}{l l l l}
		\hline
		Parameters & Description & Value and unit&Reference\\
		\hline
		TIV model& & & \cite{baccam2006kinetics}\\
		$T_k(0)$& initial number of target cells & $4\times 10^8$ cells\\
		$\alpha$& rate of infection of target cells by virus & $3.2\times 10^{-5}$ [(TCID$_{50}$/ml)$^{-1}\cdot$ d$^{-1}$]\\
		$\delta_L$& death rate of infected cells& $5.2$ d$^{-1}$\\
		$p$ & rate of increase of viral titer per infected cell & $4.6\times 10^{-2}$ [(TCID$_{50}$/ml)$^{-1}\cdot$ d$^{-1}$] \\
		$c$ & viral clearance rate& $5.2$ d$^{-1}$\\
		\hline
		Between-host model& & & Refer to text\\
		$\beta_{\rm{max}}$ & maximum contagiousness of the infected agents & $0.6$\\
		$V_{50}$ & the viral load at which contagiousness is half maximal& $10^3$ (TCID$_{50}$/ml)\\
		$\xi$ & the steepness from low to high contagiousness & 2\\
		$\epsilon$ & threshold to resign agent from infected to removed& $10^{-2}$ (TCID$_{50}$/ml)\\
		$N$ & total population size& $20000$\\
		$I(0)$ & initial number of infected agents & $10$\\
		$\lambda$ & average number of contacts of an infected agent& $1$ d$^{-1}$\\
		$t_{\rm{int}}$ & time step & $1$ hr\\
		\hline
	\end{tabular}
	\caption{Parameters and values used in the methods.}
	\label{tab:table1}
\end{table}

\section{Results}

\subsection{Multi-scale dynamics}

Classifying agents according to the phase of their viral load trajectory, as described in Section~\ref{ssec:multiscale} above, provides a unique perspective on how the viral load distribution of a population evolves over the course of an outbreak.

\begin{figure}[!ht]
	\centering
	\includegraphics[width=1 \linewidth]{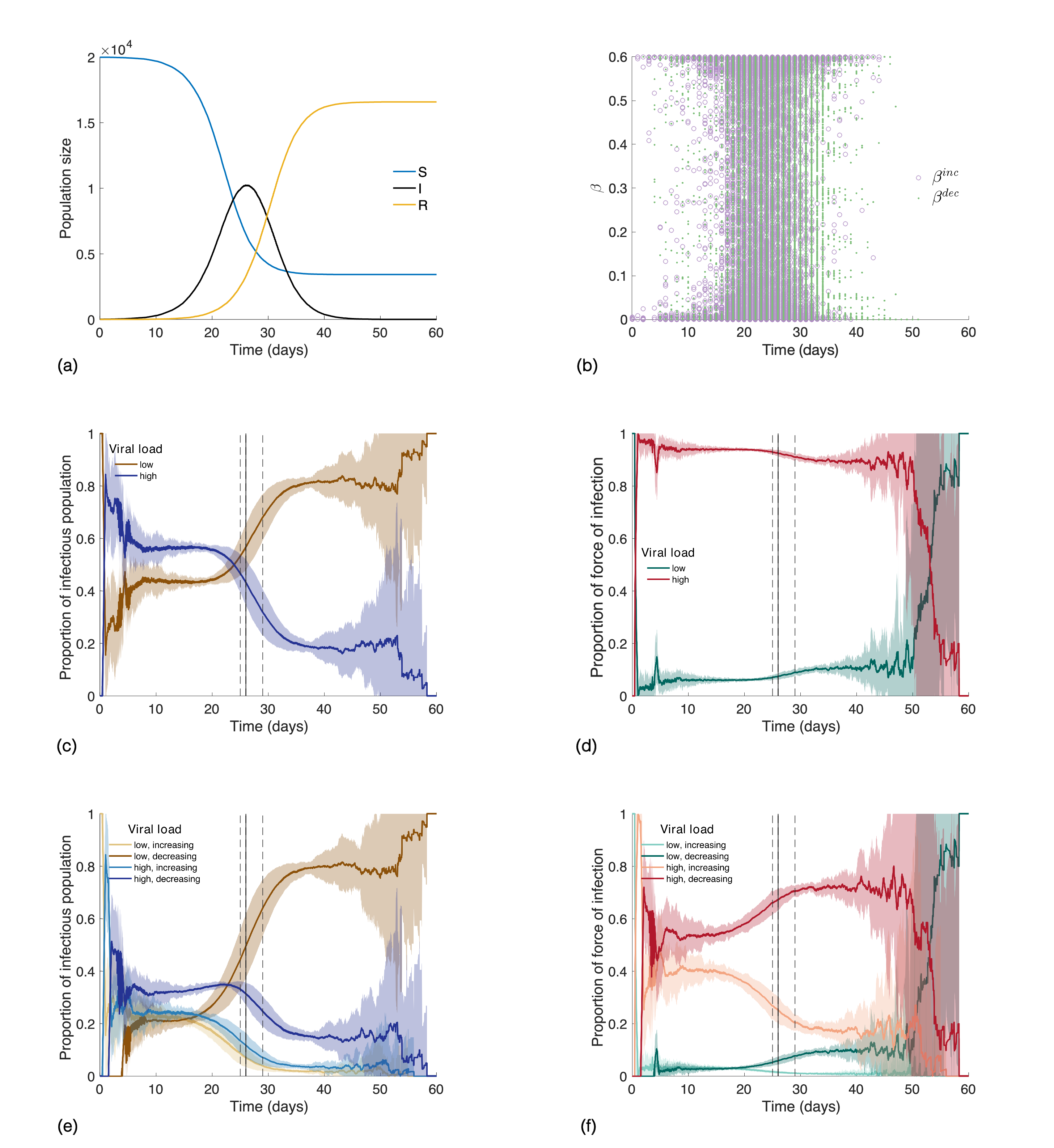}
	\caption{{\bf The distribution of infected agents with different viral load levels throughout an outbreak}. (a)~The number of agents classified as susceptible, infectious and recovered during a single exemplar simulation of the multi-scale model showing SIR-type dynamics; (b)~The contagiousness ($\beta$) of each infected agent on each day for the same simulation as in~(a); (c)--(f)~Average values over 20 simulations of the break down of the infected population on each day, by low/high viral load (c) and low-increasing/high-increasing/high-decreasing/low-decreasing viral load (e), and of the fractional contribution of those agents to the force of infection ((d) and (f) respectively). The solid and vertical dashed lines are the median and $95\%$ credible interval for the day of peak for the 20 simulations. Note that early (before day 5) and late (after day 50) in the epidemic, the small size of the infectious population leads to high levels of stochastic noise in these metrics}
	\label{fig:fast-TIV}
\end{figure}

Figure~\ref{fig:fast-TIV}(a) shows an exemplar realisation of the model displaying SIR-type epidemic dynamics. For the same realisation, Figure~\ref{fig:fast-TIV}(b) shows the contagiousness ($\beta$) of each infectious agent on each day, differentiated by whether the agent's contagious is increasing (purple circles, labeled $\beta^{inc}$) or decreasing (green circles, labeled $\beta^{dec}$) on that day. Prior to the peak (around day~27), many of the infected agents are in the increasing phase of their contagiousness, whereas following the peak, most agents are in the decreasing phase of their contagiousness. Probability (PDF) and cumulative (CDF) density functions for the distribution of infected agents' contagiousness are shown in \nameref{S3_Fig}. 

Figures~\ref{fig:fast-TIV}(c)--(f) show key aspects of our multi-scale model averaged over 20 realisations. We break the infectious population down into those with low (yellow) or high (blue) viral load (Figure~\ref{fig:fast-TIV}(c)), and further, as per Figure~\ref{fig:basic}, into those with a low increasing, high increasing, high decreasing and low decreasing viral load (Figure~\ref{fig:fast-TIV}(e)). The contribution of these subsets of the infectious population to the overall force of infection is shown in Figures~\ref{fig:fast-TIV}(d) and (f). While the proportion of infected agents with a low viral load increases throughout the outbreak, their contribution to the force of infection remains small (under 20\%). Furthermore, following the epidemic peak, the vast majority of those with a low viral load are in the decreasing phase of their infection (i.e., their contagiousness is low and decreasing), suggesting that an \emph{Adaptive}$_n$ strategy---in which those individuals with a low viral load at the time of detection are not be required to isolate---may prove an effective and efficient policy option.

\subsection{Effectiveness and efficiency of alternative testing--isolation strategies}

Assuming that routinely collected surveillance data is sufficiently available to enable real-time decision making about the need for a case to isolate, we compare the anticipated impact of alternative isolation strategies based on this information.

Figure~\ref{fig:fast-TIV_Policy} shows incidence (left panels) and cumulative incidence infection (right panels) curves for a single exemplar simulation of the agent-based model under the four strategies as defined in the Methods ({\emph{No isolation}, \emph{Isolate high}, \emph{Isolate all} and \emph{Adaptive}$_0$ with the strategy switch occurring at the peak day (i.e., $n=0$)). Results are shown under a multiple test scenario (top) and single test scenario (bottom). 
	
	Under all testing--isolation strategies, for both multiple test and a single test scenarios, isolation has a substantial effect on epidemic dynamics. Unsurprisingly, allowing for multiple tests (top rows) results in a greater reduction in transmission, and the difference between an \emph{Isolate all} (green) and \emph{Isolate high} strategy is more pronounced. Implementation of the \emph{Adaptive}$_0$ strategy (light green) results in an intermediate level of transmission, most clearly seen in the cumulative incidence curves (right panels).
	
	\begin{figure}[h]
		\centering    
		\includegraphics[width=1 \linewidth]{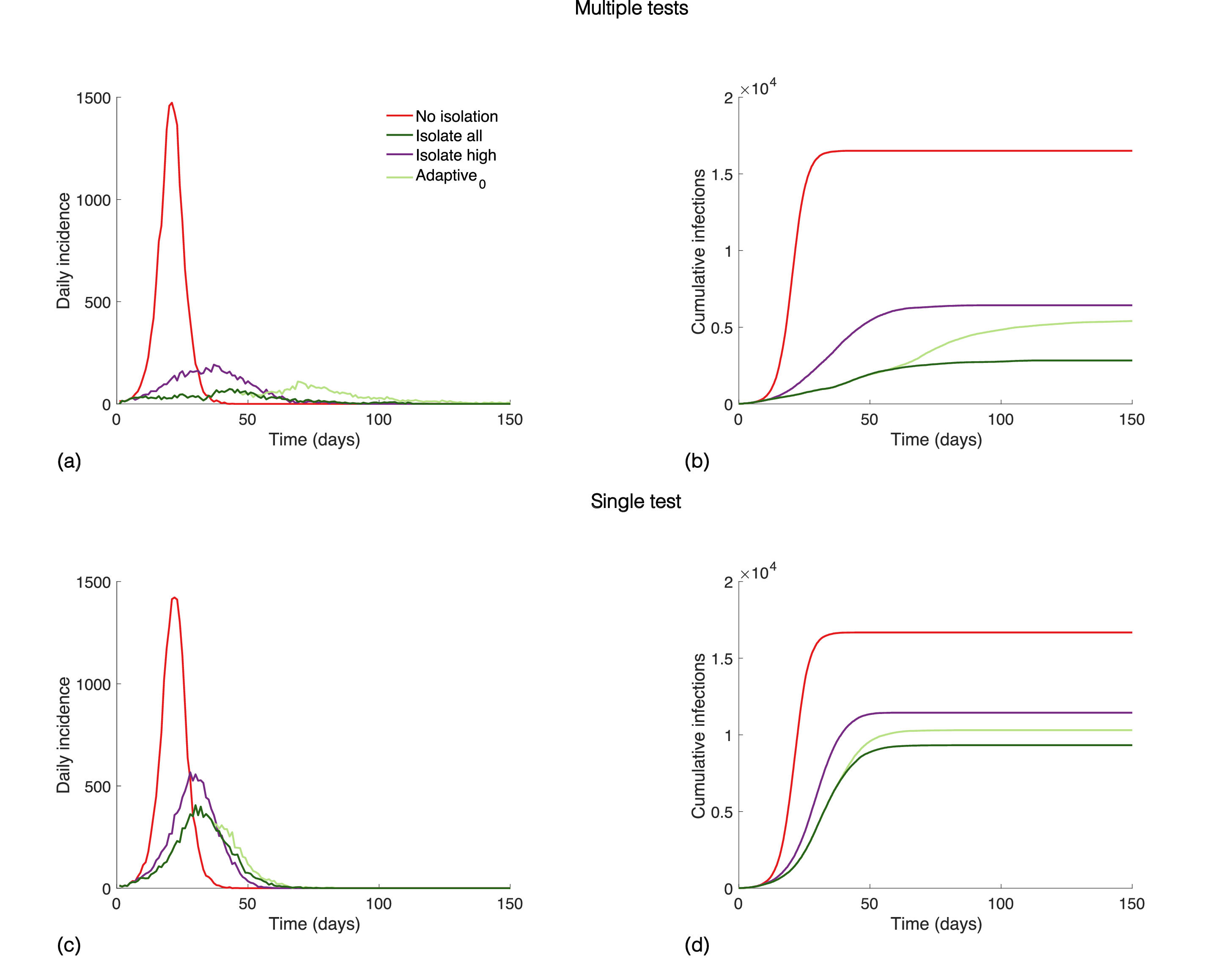}
		\caption{{\bf The impact of alternative isolation strategies on daily incidence and the final size of the epidemic under multiple and single test scenarios, for an exemplar simulation}. Here, the strategy switch in the \emph{Adaptive}$_0$ strategy occurs at the peak day. (a) and (b): Incidence and cumulative incidence for the four isolation strategies when infectious individuals may have multiple tests; (c) and (d): When infectious individuals may have only one test, regardless of outcome (positive or negative).  All isolation strategies reduce transmission, more so when multiple tests are allowed. The \emph{Adaptive}$_0$ strategy results in more infections than under the \emph{Isolate all} strategy, but a benefit of isolating those with low viral load in the lead up to the peak is clearly evident.} 
		\label{fig:fast-TIV_Policy}
	\end{figure}
	
	To evaluate the effectiveness and efficiency of alternative isolation strategies, we ran the model 100 times under each of the four strategies (no intervention and three isolation strategies), and computed their effectiveness (Equation~\ref{eq:Effectiveness}) and efficiency (Equation~\ref{eq:Efficiency}). Figure~\ref{fig:Metric} presents kernel density estimates (as violin plots) for these two metrics under multiple and single testing scenarios. Cumulative infections curves for all 100 simulations are shown in \nameref{S1_Appendix}.
	
	As expected, the \emph{Isolate all} strategy is the most effective strategy as all agents that test positive are isolated. Allowing for multiple tests results in nearly twice the number of cases averted compared to if only single tests are allowed (Figure~\ref{fig:Metric}(a) and (b)). The \emph{Adaptive}$_0$ strategy (with the switch enacted at the peak day) consistently yielded an intermediate effectiveness suggesting that, while the vast majority of those with a low viral load are in the decreasing phase of their infection following the peak, their collective contribution to transmission remains important. A more surprising result is that the efficiency---defined as the number of cases averted per person isolated---is also highest for the \emph{Isolate all} strategy and lowest for the \emph{Isolate high} strategy. Furthermore, efficiency is higher under the multiple tests scenario compared to the single test scenario (Figure~\ref{fig:Metric}((c) and (d)).
	
	\begin{figure}[!ht]
		\centering   
		\includegraphics[width=1 \linewidth]{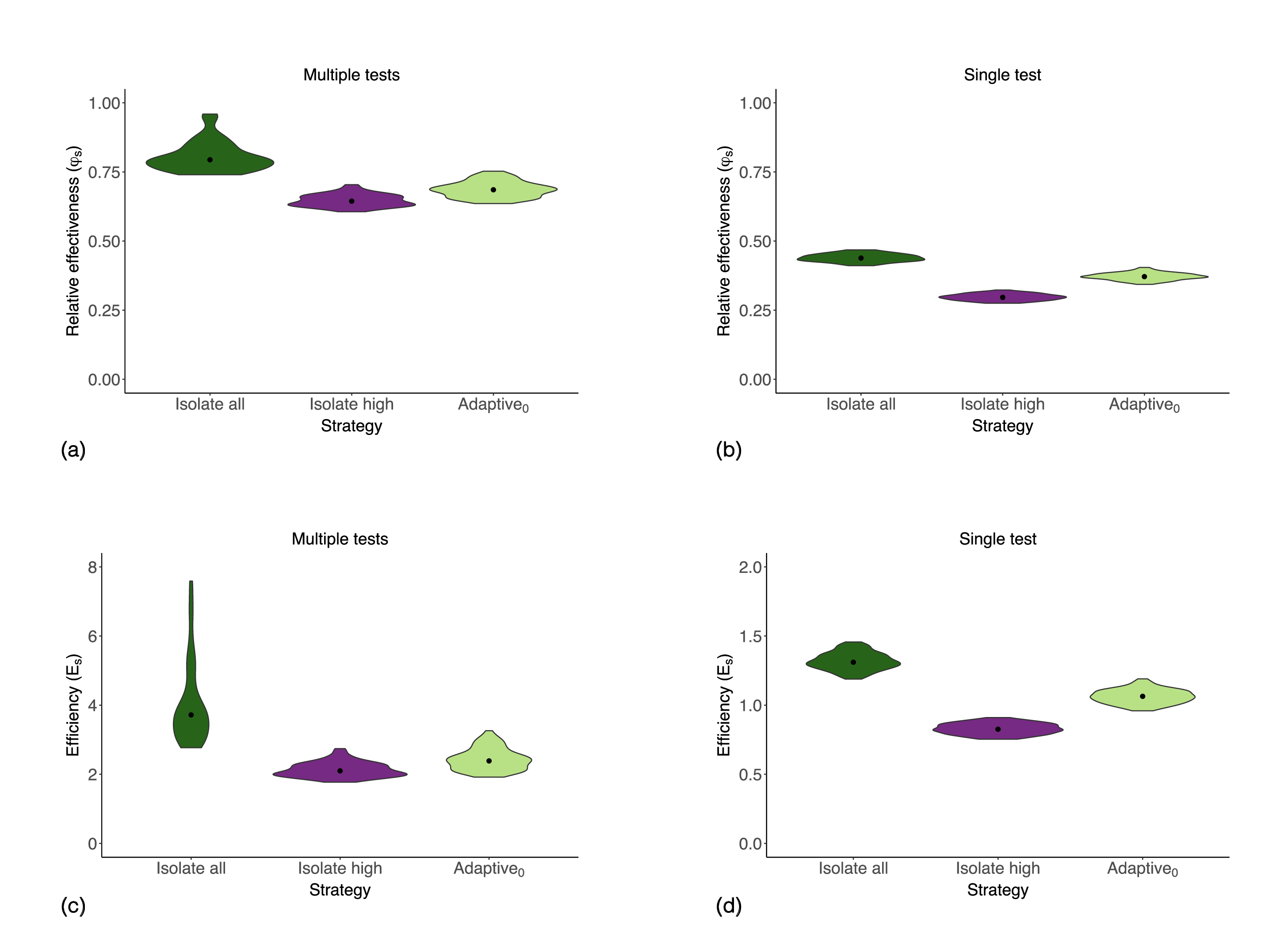}
		\caption{{\bf The effectiveness and efficiency of alternative isolation strategies under multiple and single test scenarios.} 
			Here, the strategy switch in the \emph{Adaptive}$_0$ strategy occurs at the peak day. Kernel density estimates (shown as violin plots) for the distributions (from 100 simulations of the model) of the (a)--(b) effectiveness; and (c)--(d) efficiency, of the alternative isolation strategies,  
			under multiple (panels (a) and (c)) and single (panels (b) and (d)) testing scenarios.  The \emph{Isolate all} strategy is the most effective and efficient strategy in both the multiple and single test scenarios. The \emph{Isolate high} strategy is the least effective and least efficient strategy in both the multiple and single test scenarios.
	}
	\label{fig:Metric}
\end{figure}

While enacting the switch in the \emph{Adaptive}$_0$ strategy at the peak day results in an increase in transmission and loss of efficiency compared to under an \emph{Isolate all} strategy, it may be that delaying the strategy switch (i.e., considering the \emph{Adaptive}$_n$ strategy with $n>0$) could provide better outcomes. The proportionate loss ($\psi_n$, Equation~\ref{eq:CompareSwitch}) provides a measure of this cost of switching. Figure~\ref{fig:Flu_diffday} explores how this measure of the cost of switching and the efficiency of the \emph{Adaptive}$_n$ strategy varies with the time of transition $n$ (relative to the peak day under the \emph{Isolate all} strategy) ranging from up to 5 weeks before ($n=-5$) to up to 5 weeks past ($n=5$) the peak day.

As expected, we find that switching earlier results in \emph{Isolate high}-like dynamics, while switching later results in \emph{Isolate all}-like dynamics in both multiple test scenarios (panels (a)--(c)) and single test scenarios (panels (d)--(f)). 
The proportionate loss $\psi_n$ drops from a median of around one when $n=-5$ (i.e., complete loss of benefit from isolating all individuals prior to the switch, compared to isolating only those with a high viral load) to near zero when $n=5$ (i.e., near full maintenance of the benefit of the \emph{Isolate all} strategy). Under a multiple test scenario (panel (b)) there is substantial variation in the distributions for $\psi_n$, even if switching five weeks past the peak. There is also a reasonable amount of probability mass above $\psi_5 = 0.5$, indicating a chance that over 50\% of the benefit of the \emph{Isolate all} strategy compared to the \emph{Isolate high} strategy may be lost following the switch in the \emph{Adaptive}$_5$ strategy.  The distributions for $\psi_n$ are comparatively narrower in the analogous single test scenarios (panel (e)).  Panels (c) and (f) show that efficiency systematically increases as the switching time relative to the time of peak increases with, again, more variation in the distributions in the multiple test scenarios compared to the single test scenarios.

\begin{figure}[!ht]
	\centering    
	\includegraphics[width=1\linewidth]{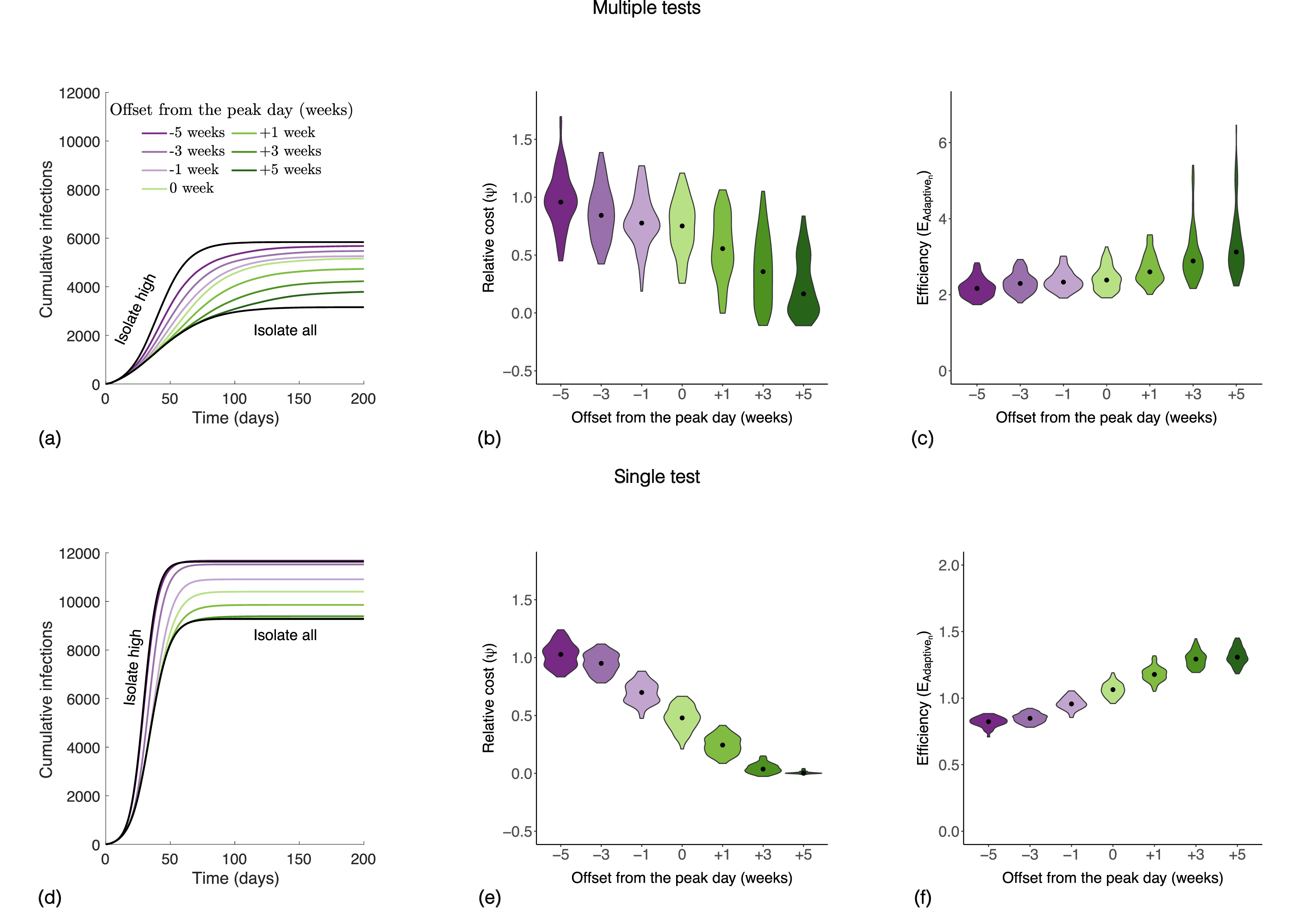}
	\caption{{\bf The impact on final size, relative cost and strategy efficiency of varying the offset in the strategy switch relative to the peak day in the \emph{Adaptive}$_n$ strategy under multiple and single test scenarios.} In all panels, the week of offset $n$ of the switch in the \emph{Adaptive}$_n$ strategy relative to the peak day is represented by the color of the lines and violin plots, where purple shading corresponds to outputs of the model run under $n<0$ scenarios, green shading to those under $n\ge0$ scenarios, and with darker shades corresponding to larger absolute values of $n$. The cumulative incidence under the \emph{Adaptive}$_n$ strategy with $n\in\{-5,-3,-1,0,1,3,5\}$ is shown in panels (a) and (d) under multiple and single test scenarios, respectively. These plots also show the cumulative incidence under the \emph{Isolate all} and \emph{Isolate high} strategies (shown in black) to illustrate how they are limiting cases of the \emph{Adaptive}$_n$ strategy when $n\to\infty$ and $n\to-\infty$, respectively. Panels (b) and (e) show the kernel density estimates of the corresponding distributions for the relative costs of the \emph{Adaptive}$_n$ strategy relative to the \emph{Isolate all} strategy ($\psi_n$) as violin plots.  Panels (c) and (f) show the kernel density estimates of the corresponding distributions for the efficiency of the \emph{Adaptive}$_n$ strategy ($E_{\text{\emph{Adaptive}$_n$}}$) as violin plots. 
	}
	\label{fig:Flu_diffday}
\end{figure}

\subsection{Sensitivity analyses: inclusion of asymptomatic infections and alternative within-host dynamics}

To this point, we have assumed that all infected agents will display symptoms and that the probability of seeking a test per unit time increases when displaying symptoms. Our particular choice of parameters for those testing probabilities means that a substantial proportion of all infections are detected as cases, with corresponding strong impact on transmission under all strategies. We next explore a scenario in which 30\% of the infected agents remain asymptomatic throughout their infectious period. Asymptomatic cases are assumed to have only a $20\%$ chance per day of being tested, fixed throughout their infectious period. This is consistent with our assumptions for pre-symptomatic infections which  have a 20\% chance of testing per day. Results are shown in \nameref{S5_Fig}. The trends in both effectiveness and efficiency across alternative strategies remain unchanged, with the obvious effect that with 30\% of the population less likely to be tested, effectiveness is lower for any given strategy.

Re-paramterising our multi-scale model with slower viral kinetics similar to SARS-CoV-2 (results presented in \nameref{S2_Appendix}) does not change any of our headline findings: an \emph{Isolate all} strategy is both most effective and the most efficient. A switch to an \emph{Adaptive}$_n$ strategy, if made post peak ($n>0$), will result in an increase in infections, but there are noticable benefits compared to isolating only those with a high viral load throughout the epidemic.

\section{Discussion}
While effective at reducing transmission, NPIs, such as case isolation, can also produce unintended health and economic costs. Therefore, it is desirable to examine pandemic response policies that would enable such measures to be targeted more selectively in order to reduce their social costs while maintaining their health benefits. Here, we explored the potential for isolation decisions based on routinely collected (during a pandemic) viral load data. Specifically, we explored the hypothesis that exempting low viral load cases throughout or only during the latter stages of an outbreak could reduce the overall number of people who are isolated with minimal effect on the size of the outbreak.

Our results suggest that a public health policy in which all cases (i.e., all detected infections) are isolated is both more effective and more efficient than a policy under which a decision to isolate is based upon an individual's viral load (as measured by a Ct value) at the time of testing. Our findings on effectiveness are unsurprising: isolating more individuals, even those with negligible future contagiousness, will of course result in fewer overall infections. However, given the changing distribution of viral loads (and so contagiousness) through time, our results on efficiency---here defined as the number of infections averted per person isolated---are noteworthy.

An adaptive strategy---in which those with a low viral load are not required to isolate, implemented at or just following the peak of an epidemic when the vast majority of those with a low viral load are in the later stage of their infection---results in a reduction in efficiency (as well as the obvious loss in effectiveness) when compared to the default strategy of isolating all cases. The adaptive strategy does, however, outperform a strategy of only isolating those with a high viral load throughout the epidemic.

Our findings are robust to alternative approaches to testing (single versus multiple tests), to parameterisations of the within-host kinetics, and to assumptions on the presence of asymptomatic infections.

The significant variation in measured effectiveness and efficiency of alternative isolation strategies, and the ``cost'' (Equation~\ref{eq:CompareSwitch}) of the adaptive strategy, is largely explained by the variation in the number of susceptible individuals remaining in the population at the time of change in strategy (\nameref{S6_Fig}). This suggests that a public health policy that sought to enact a selective isolation strategy would need to be accompanied by additional surveillance indicators, or otherwise risk poor outcomes.

We made a number of simplifying assumptions in our model. A quantitative exploration of alternative isolation strategies for a particular pathogen of interest would require incorporation of additional biological and public health policy complexity. Firstly, each agent's viral dynamics were modelled using the basic TIV model and with identical parameters (except for the initial viral load), so there is less variation across agents in our simulation than would be expected in reality. Addressing this would be technically straightforward as, at least for pathogens such as influenza and SARS-CoV-2, there are now more realistic within-host models and suitable data from which to compute population-level estimates for the joint probability distribution of within-host parameters, allowing for individual agent's viral kinetics to be sampled. While a significant undertaking, this data-driven work would enable a more realistic population to be simulated. Secondly, our testing and isolation model is very simple. We adopted simple functional forms for both the probability of seeking a test and the probability of returning a positive result given the individual's viral load. Empirical data on how both quantities relate to the time of exposure and symptom onset are well documented for some pathogens and could be incorporated into our multi-scale model if it were to be applied to a particular pathogen and response context of interest. Finally, delays to isolation following testing---which depend upon aspects of the public health laboratory reporting and surveillance system---are also important, but not modelled here. Again, empirical data on those delays could be incorporated into our flexible multi-scale framework.

A natural extension to our work would be to allow for a different (shorter) duration of isolation for those with low viral loads under the \emph{Adaptive}$_n$ strategy. It may be that a shorter period of isolation (compared to that implemented for all cases identified prior to the peak of the epidemic) would prevent the vast majority of infection events, and result in an increase in efficiency, appropriately re-defined as the number of infections averted per person-days of isolation (rather than per person isolated).

By developing a multi-scale model of transmission dynamics, we have been able to explore the potential impacts of alternative isolation strategies on epidemic dynamics. Our results indicate that even those cases with low and decreasing viral loads make a sufficient contribution to overall transmission. Any consideration of adaptive isolation policies based on viral load would require detailed evaluation using a carefully calibrated multiscale model, for which our model could provide a foundation, and likely additional surveillance streams to ensure a net positive impact of such policies.

\section{Supplementary Material}


\paragraph*{S1 Fig.}
\label{S1_Fig}
{\bf Trajectories for three exemplar agents’ contagiousness through time.}

\paragraph*{S2 Fig.}
\label{S2_Fig}
{\bf The probability (per day) for an infected agent to be tested and the probability of that test being positive as a function of the time since infection}.

\paragraph*{S3 Fig.}
\label{S3_Fig}
{\bf The probability density function (PDF) and cumulative density function (CDF) for the distribution of infected agents’ contagiousness on days 5, 15, 25, 35 and 45 for an exemplar simulation.}


\paragraph*{S4 Fig.}
\label{S5_Fig}
{\bf The relative effectiveness $\varphi_s$ and efficiency $E_s$ under \emph{Isolate all}, \emph{Isolate high} and \emph{Adaptive}$_0$ strategies when asymptomatic infections are included.}

\paragraph*{S5 Fig.}
\label{S6_Fig}
{\bf Correlation between the proportion of the population that is susceptible at the peak day and the difference in cumulative infections between the \emph{Adaptive}$_0$ and \emph{Isolate all} strategies.}


\paragraph*{S1 Appendix.}
\label{S1_Appendix}
{\bf Cumulative infections curves for all 100 simulations under multiple tests and single test scenarios.}

\paragraph*{S2 Appendix.}
\label{S2_Appendix}
{\bf The results that are applying slower viral kinetics similar to SARS-CoV-2 as the within-host model.}

\section{Acknowledgments}
We would like to thank the Australian Research Council for funding this research through Discovery Project DP210101920. JMM is supported by an ARC Laureate Fellowship (FL240100126).

\noindent This research was supported by The University of Melbourne's Research Computing Services and the Petascale Campus Initiative and an allocation on the Australian Research Data Commons NECTAR cloud computing resource.


%
%
%

\end{document}